\documentclass[11pt]{article}
\usepackage{amsmath,amsfonts,amssymb}
\usepackage{latexsym}
\usepackage{dcolumn}
\usepackage{graphicx,epsfig}
\usepackage{amsthm}
\usepackage{color}
\begin{document}
\setcounter{page}{1}

\pagestyle{plain} \vspace{1cm}
\begin{center}
\Large{\bf Unimodular $f(R,T)$ Gravity}\\
\small \vspace{1cm}{\bf Fateme
Rajabi\footnote{fa.rajabi@stu.umz.ac.ir}}\quad and\quad {\bf Kourosh
Nozari\footnote{knozari@umz.ac.ir}}\quad
 \\
\vspace{0.25cm}
Department of Physics, Faculty of Basic Sciences,\\
University of Mazandaran,\\
P. O. Box 47416-95447, Babolsar, IRAN
\end{center}

\vspace{1cm}
\begin{abstract}
We extend the idea of unimodular gravity to the modified $f(R,T)$ theories.
A new class of cosmological solutions, that the unimodular constraint on the metric imposes on the $f(R,T)$ theories, are studied.
This extension is done in both Jordan and Einstein frames. We show that while the Lagrange multiplier (that imposes the unimodular
constraint on the action) depends on the cosmic time in Jordan frame and therefore, can act as an evolving scalar field in the universe history, in the Einstein frame it acts as a cosmological constant. Then a general reconstruction method is used to realize
an explicit form of the unimodular $f(R,T)$ corresponding
to a given cosmological solution. By adopting a specific form of
$f(R,T)$, the issue of cosmological inflation is studied in this setup. To see the observational
viability of this model, a numerical analysis on the model parameter space is done in the background of Planck2015 observational data.

\end{abstract}
\newpage

\section{Introduction}

Observational data have confirmed that
our universe is currently undergoing a positively accelerating phase
of expansion~\cite{Ries98,Per99,Ber00,Han00,Pee03,Pad03}. Nowadays considering a cosmological constant (as a candidate for ``Dark Energy"
component) in the Einstein's field equations is the simplest way to
explain the late time accelerating expansion. However, suffering
from some theoretical and phenomenological problems such as the unknown origin, lake of
dynamics and also requiring a huge amount of fine-tuning for cosmological
constant's magnitude persuade the cosmologist to seek for other ways
to explain the late time acceleration. In this regard, one way is to include some sorts of scalar fields as dark energy component and as driver of late time cosmic speed up ~\cite{Ratra,Wetterich,Caldwell,Caldwell1,Caldwell2,Padmanabhan1,Sen,Sen1,Nozari2,Nozari2b}. From a geometric viewpoint, modification of the geometric part of the Einstein's field equations provides another fascinating alternative to explain large scale speed up. Braneworld models and modified gravity theories are among this alternative.
One of the simplest model of modified gravity is $f(R)$ gravity where $f$ is a generic function of the Ricci scalar,
$R$~\cite{Sot10,Noj10,Tsu08,Sta07,ADe10,Noj17, Son07,Wan10}. This model was
firstly studied by Starobinsky~\cite{Sta80} to build up a feasible
inflationary model where geometric degree of freedom has the role
of the scalar field running the cosmic inflation and the structure
formation. $f(R)$ models can describe accelerated phase of the
cosmic expansion without necessity to introduce some sorts of exotic matter.
Another modification of the general relativity consists of an
arbitrary coupling between the Ricci scalar and matter Lagrangian
density~\cite{Ber07,Har08,Tha11,OBe08,Boe08,Far09,THa10,Noj04,All05,Par08,Far07,Ber08,Har10} in the spirit of general scalar-tensor theories.
Among these non-minimally coupled theories explicit coupling of an arbitrary function of the Ricci scalar to
the stress-energy tensor of matter part has attracted attention. In this theory the gravitational
Lagrangian consists of a generic function $f(R,T)$ where $T$ is the trace of the matter stress-energy tensor.

Recently, the idea of unimodular gravity has attracted much
attention~\cite{Ant85,Wei89,Hen89,Unr89,JNg91,Fin01,Alv05,Alv07,Uza11,Jai12,Bar14,Bur15,Alv15,Sha09,Smo09,Kar12,Sal14,Alv13,Eic13,Boc07,Noz2017}.
This theory was proposed by Einstein in 1919~\cite{Ein19} without recourse to Lagrangian formalism. The idea was put into Lagrangian formalism by Anderson and
Finkelstein~\cite{And71}. The basic idea of unimodular theory is that the
determinant of the metric $\sqrt{-g}$ is fixed to a constant number
or a function of spatial coordinates. One important motivation for introduction of unimodular theory of
gravity is to address the cosmological constant problem. In this
viewpoint, the cosmological constant originates from the trace-free
part of the Einstein's field equations and not as an input parameter
into the gravitational field equations. Indeed, in this approach the
cosmological constant arises in the theory as an integration
constant. The important result of this theory is that it has the potential to
suppress the large contribution of the vacuum energy density and
cancels out the huge discrepancy between the theoretical prediction and
the observed value of the cosmological constant~\cite{Wei89}. In
this regard, the unimodular gravity can describe inflationary era
~\cite{Cho15,Bam16} as well as the late time cosmic speed up
~\cite{Jai12}. Some authors have extended this theory to
modified gravity such as $f(R)$ gravity
~\cite{Noj16,SNo16,Oik16,Odi16,Sae16} and $f({\cal{T}})$ (modified teleparallel gravity)
~\cite{Nas16,Bam16}. The authors of Ref.~\cite{Noj16} have shown that within the framework
of reconstruction method it is possible to realize various cosmological scenarios
which were impossible in standard unimodular gravity. This is the main motivation for
extension of the unimodular idea to more complicated and general situations.
We note that, it has been shown that the cosmological
perturbations of the comoving curvature perturbation, originating
from primordial quantum fluctuations in unimodular gravity are the
same as the one in ordinary general relativity at least in linear
perturbation level~\cite{Bas16,Gao14}.

In this paper we extend the idea of unimodular gravity to $f(R,T)$ modified
gravity. Our motivation is to explore some yet unknown aspects of cosmological
solutions in the spirit of $f(R,T)$ theories. In other words,
we show that a unimodular extension of $f(R,T)$ theories reveals some
new features that were impossible to be realized in standard $f(R,T)$ theories.
Since the Friedmann-Robertson-Walker metric does not
satisfy the unimodular constraint, following~\cite{Noj16} we introduce a new time variable
and then we derive the gravitational equations in Jordan frame within the metric formalism.
As an important point, we show that in the Jordan frame the Lagrange multiplier, $\lambda$, depends on the cosmic time.
So, this multiplier can act as a scalar field in the cosmic evolution. This may be a way to shed light on the dark energy problem
and can lead to the late time cosmic accelerated expansion. However, this Lagrange multiplier is constant
in Einstein frame. Hence, the  emerging constant $\lambda$ in this frame would provide the accelerating stage of the universe evolution.
Unlike the Jordan frame, the determinant of the metric in the Einstein frame is not a constant.

We use the reconstruction method to realize an explicit form of
unimodular $f(R,T)$ corresponding to a given cosmological
solution. We also study the cosmological inflation in this unimodular $f(R,T)$ model in
the Jordan frame. We calculate some inflation quantities such as the
slow-roll parameters, the scalar spectral index, its running, the tensor
spectral index and the tensor-to-scalar ratio that help us to a
better understanding of the theory and its cosmological viability~\cite{Lid94,Lid97,Gor11,Mar14,Ade14,PAde16}.
After that, we transform to the Einstein frame using a conformal
transformation and derive the gravitational equations in this
frame. We study cosmological inflation in Einstein frame too.
Finally we compare our results with Planck 2015 TT, TE, EE+low P data for some specific $f(R,T)$.

This paper is organized as follows: In section 2, we
present a brief review of the unimodular gravity to introduce the main idea and notations. We explain how
a cosmological constant is obtained without adding it by hand in
the standard Einstein-Hilbert action. In section 3 we introduce the unimodular $f(R,T)$ gravity for the first time and
derive the field equations on the unimodular FRW background. In
section 4, we consider some specific types of the scale factor and
through a reconstruction method we investigate which specific
unimodular $f(R,T)$ model can realize these cosmologies. In section 5
we study the issue of cosmological inflation in the context of unimodular $f(R,T)$ gravity. We calculate the
slow-roll and inflation parameters with a specific type of $f(R,T)$
and we compare our results with the Planck2015 observational data. In section 6, the
conformal transformation to Einstein frame is performed and the
corresponding unimodular version is obtained in
Einstein frame. Section 7 is devoted to a summary and conclusion.\\

\section{Unimodular Gravity}

The basic idea of unimodular gravity is that the determinant of the
spacetime metric is not dynamical, whereas the components of the
metric are dynamical. This means that the determinant of the metric
is fixed as
\begin{equation}
\sqrt{-g}=\epsilon_{0}\,,
\end{equation}
where $\epsilon_0$ is a constant parameter. We can impose this
constraint by including the Lagrange multiplier in the action as
follows
\begin{equation}
S=\frac{1}{2\kappa^2}{\int}{d^4x\Big[\sqrt{-g}R-2\lambda(\sqrt{-g}-\epsilon_{0})\Big]}
+S_{m}\,,
\end{equation}
where $R$ is the Ricci scalar of the physical metric $g_{\mu\nu}$,
$\lambda$ is the Lagrange multiplier which basically is dynamical,
and $\kappa^2=8\pi G$. Also, $S_m$ is the matter field's action.
Indeed, we can get unimodular constraint (1) by varying the action (2)
with respect to parameter $\lambda$. Moreover, the variation of the action
with respect to the metric gives the following Einstein's field
equations
\begin{equation}
R_{\mu\nu}-\frac{1}{2}g_{\mu\nu}R+g_{\mu\nu}\lambda=\kappa^2
T_{\mu\nu}\,.
\end{equation}
In the standard Einstein gravity there are $10$ independent field equations. However, in unimodular gravity there are
only $9$ independent components because of the extra constraint on the
determinant of the metric. The field equations (3) are
the same as the Einstein's field equation with a cosmological
constant. By taking the divergence of the field equations we get
\begin{equation}
\nabla^{\mu}\Big(R_{\mu\nu}-\frac{1}{2}g_{\mu\nu}R+g_{\mu\nu}\lambda-\kappa^2
T_{\mu\nu}\Big)=0\,.
\end{equation}
Then, by using the Bianchi identities,
$\nabla^{\mu}(R_{\mu\nu}-\frac{1}{2}g_{\mu\nu}R)=0$, and the
conservation of the energy-momentum tensor,
$\nabla^{\mu}T_{\mu\nu}=0$, we get
\begin{equation}
\nabla^{\mu}\lambda=0 \,\,\rightarrow \,\, \lambda=\lambda_0\,.
\end{equation}
The value of $\lambda_0$ is obtained by taking the trace of the
field equations (3) as
\begin{equation}
R+\kappa^2 T=4\lambda_0\,.
\end{equation}
In this regard, given that in action (2) $\lambda_0$ is a constant
parameter, the usual Einstein's field equations are reproduced and
$\lambda_0$ plays the role of a cosmological constant. This is the
main importance of the unimodular gravity. Classically, the
unimodular gravity recovers the same physics as the general
relativity with a cosmological constant. The difference is that the
cosmological constant appears as an integration constant in the
unimodular gravity which can take any value. In this regard, this
constant can fix the value of the vacuum energy density.

\section{Unimodular $f(R,T)$ Gravity in Jordan Frame}

In this section, we generalize the unimodular Einstein-Hilbert
gravity to the $f(R,T)$ modified theory of gravity with $T$ being
the trace of the stress-energy tensor, $T_{\mu\nu}$.  As we have stated previously,
we show that a unimodular extension of $f(R,T)$ gravity reveals some
new features in the spirit of cosmology that were impossible to be realized in the standard $f(R,T)$ theories.
We work in Jordan frame and within the metric formalism. The action of
the unimodular $f(R,T)$ gravity in Jordan frame is given by
\begin{equation}
S=\frac{1}{2\kappa^2}{\int}{d^4x\Big[\sqrt{-g}f(R,T)-2\lambda(\sqrt{-g}-\epsilon_{0})\Big]}
+{\int}d^4x\sqrt{-g}{\cal{L}}_{m}\,\,,
\end{equation}
where ${\cal{L}}_{m}$ is the matter Lagrangian density. The stress-energy
tensor of the matter fields is defined as
\begin{equation}
T_{\mu\nu}=-\frac{2}{\sqrt{-g}}\frac{\delta(\sqrt{-g}{\cal{L}}_{m})}{\delta
g^{\mu\nu}}=g_{\mu\nu}{\cal{L}}_{m}-2\frac{\partial{\cal{L}}_{m}}{\partial
g^{\mu\nu}}\,.
\end{equation}
By varying action (7) with respect to the metric, we obtain the
following modified Einstein's field equations
\begin{eqnarray}
f_{,R}(R,T)R_{\mu\nu}-\frac{1}{2}g_{\mu\nu}f(R,T)+(g_{\mu\nu}\Box
-\nabla_{\mu}\nabla_{\nu})f_{,R}(R,T)+\lambda g_{\mu\nu}=\kappa^2
T_{\mu\nu}-f_{,T}(R,T)(T_{\mu\nu}+\Theta_{\mu\nu})\,.
\end{eqnarray}
where $f_{,X}\equiv\frac{df}{dX}$. In this relation $\Theta_{\mu\nu}$ is defined as
\begin{equation}
\Theta_{\mu\nu}\equiv g^{\alpha\beta} \frac{\delta
T_{\alpha\beta}}{\delta g^{\mu\nu}}\,.
\end{equation}
Variation of the action (7) with respect to $\lambda$ gives the
unimodular condition $\sqrt{-g}=\epsilon_{0}$. Taking the trace of
the field equations (9) leads to
\begin{equation}
f_{,R}(R,T)R+3\Box f_{,R}(R,T)-2f(R,T)+4\lambda=\kappa^2
T-f_{,T}(R,T)(T+\Theta)\,.
\end{equation}
By using this relation we rewrite the field equations (9) as follows
\begin{eqnarray}
&f_{,R}(R,T)\Big[R_{\mu\nu}-\frac{1}{3}g_{\mu\nu}R\Big]
-\nabla_{\mu}\nabla_{\nu}f_{,R}(R,T)+\frac{1}{6}
f(R,T)g_{\mu\nu}-\frac{1}{3}\lambda
g_{\mu\nu}=\nonumber\\
&\kappa^2\Big[T_{\mu\nu}-\frac{1}{3}g_{\mu\nu}T\Big]
-f_{,T}(R,T)\Big[T_{\mu\nu}-\frac{1}{3}g_{\mu\nu}T\Big]
-f_{,T}(R,T)\Big[\Theta_{\mu\nu}-\frac{1}{3}g_{\mu\nu}\Theta\Big]\,.
\end{eqnarray}
In this regard, the standard $f(R,T)$ equations as derived in Ref.~\cite{Har11} are recovered with an
additional cosmological constant. The covariant derivative of the
field equations (9) gives
\begin{eqnarray}
\nabla^{\mu}T_{\mu\nu}=\frac{1}{\kappa^2-f_{,T}(R,T)}
\Big[-\frac{1}{2}g_{\mu\nu}f_{,T}(R,T)\nabla^{\mu}T+(T_{\mu\nu}
+\Theta_{\mu\nu})\nabla^{\mu}f_{,T}(R,T)\\ \nonumber+f_{,T}(R,T)\nabla^{\mu}\Theta_{\mu\nu}
+g_{\mu\nu}\nabla^{\mu}\lambda\Big]\,,
\end{eqnarray}
where we have used the relation
$(\nabla_{\mu}\Box-\Box\nabla_{\mu})f_{,R}=R_{\mu\nu}\nabla^{\nu}f_{,R}$.
Equations (13) imply that the stress-energy tensor of the matter
fields is not conserved and this is due to the interaction between
the matter and curvature sectors. By using equations (8) and (10), we obtain
$\Theta_{\mu\nu}$ as
\begin{equation}
\Theta_{\mu\nu}=-2T_{\mu\nu}+g_{\mu\nu}{\cal{L}}_{m}
-2g^{\alpha\beta}\frac{{\partial^2}{\cal{L}}_{m}}{\partial
g^{\mu\nu}\partial g^{\alpha\beta}}\,.
\end{equation}
In this paper, we assume a perfect fluid to be the
source of the stress-energy tensor with
\begin{equation}
T_{\mu\nu}=(\rho_m+p_m)u_{\mu}u_{\nu}+p_m g_{\mu\nu}\,,
\end{equation}
where $u_{\mu}$ is the four-velocity with $u_{\mu}u^{\mu}=-1$.
$\rho_m$ and $p_m$ are the energy density and pressure of the matter
fields we assume to be related as $p_m={\omega_m}\rho_m$, where $\omega_m$ is the
equation of state parameter. In this case, by comparing equations (8) and (15) the matter Lagrangian can
be set as ${\cal{L}}_{m}=p_m$ (see [66] for details). Therefore, we have
$\Theta_{\mu\nu}$ as
\begin{equation}
\Theta_{\mu\nu}=-2T_{\mu\nu}+p_m g_{\mu\nu}\,.
\end{equation}
Now, the field equations (9) can be written as
\begin{eqnarray}
f_{,R}(R,T)R_{\mu\nu}-\frac{1}{2}g_{\mu\nu}f(R,T)+\big(g_{\mu\nu}\Box-\nabla_{\mu}\nabla_{\nu}\big)f_{,R}(R,T)
+\lambda g_{\mu\nu}=\kappa^2 T_{\mu\nu}\\ \nonumber
+f_{,T}(R,T)
T_{\mu\nu} -p_m g_{\mu\nu}f_{,T}(R,T)\,.
\end{eqnarray}
This equation can be rewritten as
\begin{equation}
G_{\mu\nu}\equiv R_{\mu\nu}-\frac{1}{2}g_{\mu\nu}R=8\pi
G_{eff}T_{\mu\nu}+T_{\mu\nu}^{(eff)}\,,
\end{equation}
where
\begin{equation}
G_{eff}\equiv\frac{1}{f_{,R}}G\Big(1+\frac{f_{,T}}{8\pi G}\Big),
\end{equation}
\begin{equation}
T_{\mu\nu}^{(eff)}\equiv\frac{1}{f_{,R}}\Big[\frac{1}{2}g_{\mu\nu}
(f-Rf_{,R}-2p_m
f_{,T}-2\lambda)-(g_{\mu\nu}\Box-\nabla_{\mu}\nabla_{\nu})f_{,R}\Big]\,\,.
\end{equation}
In this relation, the quantity $G_{eff}$ can be regarded as an effective
gravitational coupling strength which depends on the derivatives of
$f(R,T)$. $T_{\mu\nu}^{(eff)}$ is an effective stress-energy tensor
which includes both the geometry and matter contributions simultaneously. In this
approach, corrections are applied to the right hand side of
the Einstein's field equations. So, one can say that the cosmic
speed up results in from a geometrical contribution to the total
cosmic energy density and the matter content of the universe simultaneously.
To be more clarified, in order to compare equation (17) with Einstein's equations,
we have rewritten the field equations (17) in the form of effective Einstein field equations, (18).
As we have stated, in this form $T_{\mu\nu}$ is the stress-energy of the standard matter which is assumed to be a
perfect fluid and $G_{eff}$ plays the role of an effective gravitational coupling.
It is obvious that for the case of general relativity, $f(R,T)=R$ and $G_{eff}=G$
which means that the effective gravitational constant reduces to the standard Newtonian gravitational constant.
But, for the second term on the right hand side of Eq. (18), following the standard literature such as Ref.~\cite{Har11},
$T_{\mu\nu}^{eff}$ is an effective stress-energy tensor which is dependent on the geometry and matter
contributions simultaneously. It contains the contribution of modification of the geometric part of
the theory and therefore in a dark energy perspective, this part is responsible for explanation of
the late time cosmic speed up since the standard matter has not such a capability.
This is the part that is usually dubbed as the "dark fluid" or the "dark curvature" in literature.
In other words, by writing the field equations as (18) our aim was to focus on the role of the dark sector
coming from modification of the gravitational theory and unimodularity from the rest of the standard matter contribution.
We note that we can consider $8\pi G_{eff}T_{\mu\nu}$ as an effective stress-energy tensor for the standard matter.
The Bianchi identity then gets the following form $0=8\pi(\nabla^{\mu} G_{eff}) T_{\mu\nu}+8\pi G_{eff} (\nabla^{\mu}T_{\mu\nu})+\nabla^{\mu}T_{\mu\nu}^{eff}$.
By using equations (19) and (20) and also using the field equations, equation (13) is recovered.

Now, we consider a spatially flat Friedmann-Robertson-Walker (FRW)
space-time with the following metric
\begin{equation}
ds^{2}= -dt^{2} + a^{2}(t)dx_{i}dx^{i},\quad i=1,2,3
\end{equation}
where $a(t)$ is the scale factor. However, this metric does not
satisfy the unimodular constraint (1). To solve this problem, following~\cite{Noj16} we
introduce a new time variable as
\begin{equation}
d\tau=a^3(t)dt\,,
\end{equation}
By this definition, the FRW metric (21) can be rewritten as
\begin{equation}
ds^{2}= -a^{-6}(\tau)d\tau^{2} +a^2(\tau)dx_{i}dx^{i},\quad i=1,2,3
\end{equation}
where $g_{\mu\nu}=diag(-a^{-6}(\tau),a^2(\tau),a^2(\tau),a^2(\tau))$
with the unimodular constraint $\sqrt{-g}=1$. Now, with metric (23),
the non-vanishing components of the Ricci tensor and also the Ricci
scalar are as follows
\begin{eqnarray}
R_{\tau\tau}=-3\dot{{\cal{H}}}-12{\cal{H}}^2\,,\quad
R_{ij}=a^8(\dot{{\cal{H}}}+6{\cal{H}}^2)\,,\quad
R=a^6(6\dot{{\cal{H}}}+30{\cal{H}}^2)\,.
\end{eqnarray}
where, ${\cal{H}}$ is the Hubble parameter defined as
${\cal{H}}=\frac{1}{a}\frac{da}{d\tau}$. By using these equations
and unimodular FRW metric (23), we obtain the $\tau\tau$ and $ii$
components of the field equations as
\begin{eqnarray}
-(3\dot{{\cal{H}}}+12{\cal{H}}^2)f_{,R}+\frac{1}{2}(f-2\lambda)a^{-6}+3{\cal{H}}\dot{f}_{,R}=
[\kappa^2\rho_m+f_{,T}(\rho_m+p_m)]a^{-6}\,,
\end{eqnarray}
and
\begin{eqnarray}
(\dot{{\cal{H}}}+6{\cal{H}}^2)f_{,R}-\frac{1}{2}(f-2\lambda)a^{-6}-5{\cal{H}}\dot{f}_{,R}
-\ddot{f}_{,R}=\kappa^2p_ma^{-6}\,,
\end{eqnarray}
where, a dot represents derivative with respect to $\tau$. By
contracting equations (25) and (26) we find
\begin{eqnarray}
(2\dot{{\cal{H}}}+6{\cal{H}}^2)f_{,R}+2{\cal{H}}\dot{f}_{,R}+\ddot{f}_{,R}
=-(\kappa^2+f_{,T})(\rho_m+p_m)a^{-6}\,.
\end{eqnarray}
This equation has an important role in the reconstruction method. By
using the metric (23) and equation (13) we get
\begin{eqnarray}
\dot{\rho}_m+3{\cal{H}}(\rho_m+p_m)=-\frac{1}{\kappa^2
+f_{,T}}\big[-\frac{1}{2}\dot{T}f_{,T}+(\rho_m+p_m)
\dot{f}_{,T}+\dot{p}_{m}f_{,T}+\dot{\lambda}\big]\,.
\end{eqnarray}
This equation shows that a general unimodular $f(R,T)$ model does not
satisfy the usual conservation law. By assuming that
$T_{\mu\nu}^{(eff)}$ behaves as the perfect fluid, equation (20) gives
the following effective energy density and pressure
\begin{equation}
\rho_{eff}=\frac{1}{f_{,R}}\big[-\frac{1}{2}(f-Rf_{,R}-2p_mf_{,T}
-2\lambda)-3a^6{\cal{H}}\dot{f}_{,R}\big],
\end{equation}
\begin{equation}
p_{eff}=\frac{1}{f_{,R}}\big[\frac{1}{2}(f-Rf_{,R}-2p_m
f_{,T}-2\lambda)+a^6(\ddot{f}_{,R}+5{\cal{H}}\dot{f}_{,R})\big],
\end{equation}
respectively. The equation of state parameter in this model is
obtained as
\begin{equation}
\omega_{eff}=\frac{p_{eff}}{\rho_{eff}}=-1+\frac{a^6(\ddot{f}_{,R}+2{\cal{H}}\dot{f}_{,R})}
{-\frac{1}{2}(f-Rf_{,R}-2p_{m}f_{,T}-2\lambda)-3a^6{\cal{H}}\dot{f}_{,R}}
\end{equation}
which depends on $f(R,T)$ and its derivatives. Figure $1$ shows the
behavior of the effective equation of state parameter $w_{eff}$
versus the redshift, $z$, for a specific $f(R,T)$ candidate. As this figure shows, the phantom divide
crossing occurs at $z\simeq0.2$.

\begin{figure}[htp]
\begin{center}\includegraphics{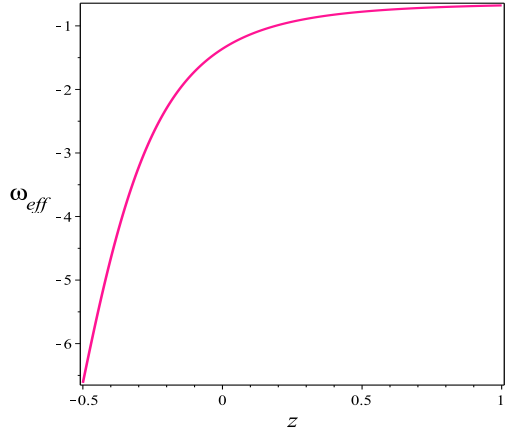} \vspace{5.6cm}
\end{center}
\caption{\small{The effective equation of state parameter versus the
redshift for $f(R,T)=R^p+\alpha R^n+\beta T^m+T$ and $\omega_m=0$.
To plot this figure we have set $p=0.49$, $n=0.65$, $m=2$ and
$\alpha=\beta=1$. Crossing of the phantom divide line occurs at
$z\simeq0.2$.}}
\end{figure}

\section{Reconstruction of the unimodular $f(R,T)$ gravity}

In this section we study reconstruction of modified gravity with unimodular $f(R,T)$
action. In the first step, we adopt the power-law scale factor as
\begin{equation}
a(t)=\left(\frac{t}{t_0}\right)^\alpha \quad \Rightarrow \qquad
H=\frac{\alpha}{t} \,,
\end{equation}
where, $t_0$ and $\alpha$ are constant parameters. This solution
is able to describe the evolution of the scale factor for standard model
universe (such as a dust dominated universe corresponding to $\alpha
= 2/3$, or a radiation dominated universe corresponding to $\alpha =
1/2$). Also, $\alpha>1$ gives an accelerating expansion. By
substituting this scale factor into equation (22) and integrating we obtain
\begin{equation}
\tau=\frac{t_0}{3\alpha+1}\Big(\frac{t}{t_0}\Big)^{3\alpha+1}\,,
\end{equation}
and substituting this equation into equation (32) we get
\begin{equation}
a(\tau)=\Big({\frac{3\alpha+1}{t_0}}\tau
\Big)^{\frac{\alpha}{3\alpha+1}} \,.
\end{equation}
Let us now consider the following exponential scale factor
(corresponding to the de Sitter universe)
\begin{equation}
a(t)=e^{{H_0}t}\,,
\end{equation}
which enables one to describe both the initial inflation and late-time
cosmic acceleration. In equation (35), $H_0$ is a positive constant. In this
case equation (22) leads to
\begin{equation}
\tau=\frac{1}{3H_0}e^{3H_0t}\,.
\end{equation}
Now, we find the scale factor of the universe in terms of the new
variable $\tau$ as follows
\begin{equation}
a(\tau)=\Big(3H_0 \tau\Big)^{\frac{1}{3}}\,.
\end{equation}
Note that in the limit $\alpha\rightarrow\infty$ and
$t_0\rightarrow\infty$, with
$\frac{3\alpha+1}{t_0}\rightarrow{3H_0}$, the power-law expansion
(34) gives the de Sitter expansion (37). Now we consider the
following form of the scale factor
\begin{equation}
a(\tau)=\Big(\frac{\tau}{\tau_0}\Big)^q \quad \Rightarrow \qquad
{\cal{H}}=\frac{q}{\tau}\,,
\end{equation}
with $q$ and $\tau_0$ being constant parameters. In this regard,
the unimodular FRW metric (23) takes the following form
\begin{equation}
ds^2=-\Big(\frac{\tau}{\tau_0}\Big)^{-6q}d\tau^2+\Big(\frac{\tau}
{\tau_0}\Big)^{2q}dx_{i}dx^{i},\qquad i=1,2,3
\end{equation}
In this general case, a de Sitter cosmological evolution occurs when
$q=\frac{1}{3}$. When $q=\frac{2}{9}$ and $q = \frac{1}{5}$, the
universe is  respectively dominated by dust and radiation.
$\frac{1}{4}<q<\frac{1}{3}$, corresponding to $\alpha>1$, shows an
accelerating universe. By using equation (38), the Ricci scalar
becomes as
\begin{equation}
R=\frac{6q(5q-1)}{\tau_0^2}\Big(\frac{\tau}{\tau_0}\Big)^{6q-2}\,.
\end{equation}
Now, as equation (13) shows, energy-momentum of the standard matter
is not covariantly conserved in this unimodular $f(R,T)$ gravity.
Therefore, a test particle that is moving in a gravitational field
as described here does not follow a geodesic line in the sense of General Relativity. In fact, the
coupling between matter and geometry in this framework induces an
extra acceleration acting on the particle. This is because of
interaction between the standard matter and the ``dark fluid" (an
effective fluid description coming from modification of the
geometric sector of the standard general relativity, as is usually
interpreted in modified gravity literature and dubbed also as
``curvature fluid"). This non-conservation of the standard matter
leads to violation of the usual global evolution (as supported by
observation) of the different species in the universe. However, the
covariant conservation of energy-momentum tensor  $T_{\mu\nu}$ of
the standard matter is an essential feature in standard general
relativity which is a direct consequence of the diffeomorphism
invariance of the theory. So, it is expected that any classical
gravitational theory should satisfy such a requirement as well to
give usual global cosmic evolution of the standard species. What we
are going to do here is that for a moment we assume (as is usual in
$f(R,T)$ literature, see for instance~\cite{Har11,Alvarenga13}) that
such a conservation to be satisfied globally and then, as a result,
we are able to find a constraint on the field equations. In other
words, we pursue in such a way that if we insist on the conservation
of energy-momentum as a consequence of diffeomorphism invariance of the theory, what will happens for the
field equations. For this purpose, we set the right hand side of
equation (28) to be zero which imposes a constraint on the model's
field equations as follows
\begin{eqnarray}
-\frac{1}{2}\dot{T}f_{,T}+(\rho_m+p_m)
\dot{f}_{,T}+\dot{p}_{m}f_{,T}+\dot{\lambda}=0
\end{eqnarray}
where $T=T^{\mu}_{\mu}=-\rho_m+3p_m$. Hence, for a perfect fluid
with an equation of state $p_m=\omega_m \rho_m$, where $\omega_m$ is
a constant, the above equation can be written as
\begin{eqnarray}
\frac{1}{2}(1-\omega_m)\dot{\rho}_mf_{,T}+(1+\omega_m)\rho_m
\dot{f}_{,T}+\dot{\lambda}=0
\end{eqnarray}
In this respect, the conserved matter contents of the universe
satisfy the relation
\begin{eqnarray}
\dot{\rho}_m=-3{\cal{H}}(1+\omega_m)\rho_m \quad \longrightarrow
\quad \rho_m=\rho_{0m} a^{-3(1+\omega_m)} \,,
\end{eqnarray}
We can write these parameters in terms of the time parameter $\tau$
as follows
\begin{eqnarray}
\dot{\rho}_m=\Big(\frac{\tau}{\tau0}\Big)^{-3(1+\omega_m)q}\quad ,
\quad T=-(1-3\omega_m)\Big(\frac{\tau}{\tau0}\Big)^{-3(1+\omega_m)q}
\,.
\end{eqnarray}

$\bullet \quad f(R,T)=f_{1}(R)+f_{2}(T)$ \vspace{0.4cm}

The first specific type of $f(R,T)$ we adopt here is defined as
$f(R,T)=f_{1}(R)+f_{2}(T)$, where $f_1(R)$ and $f_2(T)$ are
arbitrary functions of their argument. For this type of $f(R,T)$,
equation (27) get
\begin{equation}
\frac{2q(3q-1)}{\tau^2}f_{1,R}+2\frac{q}{\tau}\dot{f}_{1,R}+\ddot{f}_{1,R}+
(\kappa^2+f_{2,T})(1+\omega)(\frac{\tau}{\tau_0})^{-3(3+\omega)q}=0\,,
\end{equation}
and the equation (42) become
\begin{equation}
-\frac{3q}{2\tau}(1-\omega^2)(\frac{\tau}{\tau_0})^{-3(1+\omega)q}
f_{2,T}+(1+\omega)(\frac{\tau}{\tau_0})^{-3(1+\omega)q}\dot{f}_{2,T}+\dot{\lambda}=0\,,
\end{equation}
Hence, by combining the equations (26) and (46) and canceling
$\lambda$, we obtain the following equation
\begin{eqnarray}
\frac{(-30q^2+54q^3+4q)}{\tau^3}\,\,f_{1,R}+\frac{(-4q+24q^2)}{\tau^2}\,\,\dot{f}_{1,R}
+\frac{11q}{\tau}\,\,\ddot{f}_{1,R}\\ \nonumber
+\dddot{f}_{1,R}-3\omega_m(1+\omega_m)
(\kappa^2+f_{2,T})\frac{q}{\tau}(\frac{\tau}{\tau_0})^{-3(1+\omega_m)q}=0\,,
\end{eqnarray}
By solving the differential equations (45) and (47) we obtain
\begin{equation}
f_{1,R}(\tau)=C_+\tau^{\mu_+}+C_- \tau^{\mu_-}+A\tau^{-9q+2}\,,
\end{equation}
where $A$ and $C_{\pm}$ are integration constants and
\begin{equation}
\mu_{\pm}=\frac{-2q+1\pm\sqrt{-20q^2+4q+1}}{2}\,.
\end{equation}
By using equation (40), the solution (48) can be expressed in terms
of $R$ as follows
\begin{eqnarray}
f_{1,R}(R)=C_+{\tau_0}^{\mu_+}\Big(\frac{\tau_0^2}{30q^2-6q}\Big)^{\frac{\mu_+}
{6q-2}}R^{\frac{\mu_+}{6q-2}}+C_- {\tau_0}^{\mu_-}
\Big(\frac{\tau_0^2}{30q^2-6q}\Big)^{\frac{\mu_-}
{6q-2}}R^{\frac{\mu_-}{6q-2}}\nonumber\\
+A{\tau_0}^{-9q+2}\Big(\frac{\tau_0^2}{30q^2-6q}
\Big)^{\frac{-9q+2}{6q-2}}R^{\frac{-9q+2}{6q-2}}\,,\hspace{4.5cm}
\end{eqnarray}
By integrating of the above equation with respect to Ricci scalar,
we obtain
\begin{equation}
f_{1}(R)=B_{+}R^\frac{\mu_{+}+6q-2}{6q-2}+B_{-}R^\frac{\mu_{-}
+6q-2}{6q-2}+DR^\frac{-3q}{6q-2}\,,
\end{equation}
where
\begin{equation}
B_{\pm}=C_{\pm}{\tau_0}^{\mu_{\pm}}\Big(\frac{6q-2}{\mu_{\pm}+6q-2}\Big)\Big(\frac{\tau_0^2}
{30q^2-6q}\Big)^{\frac{\mu_\pm}{6q-2}}\,,
\end{equation}

\begin{equation}
D=-A{\tau_0}^{-9q+2}\Big(\frac{6q-2}{3q}\Big)\Big(\frac{\tau_0^2}{30q^2-6q}\Big)^{\frac{-9q+2}
{6q-2}}\,.
\end{equation}
and
\begin{equation}
f_{2,T}(\tau)=-A\Big(\frac{69q^2-25q+2}{1+\omega_m}\Big){\tau_0}^{-9q}\Big
(\frac{\tau}{\tau_0}\Big)^{3\omega_m q}-\kappa^2\,,
\end{equation}
which by using equation (44) is rewritten as
\begin{equation}
f_{2,T}(T)=-A\Big(\frac{69q^2-25q+2}{1+\omega_m}\Big){\tau_0}^{-9q}\Big
(\frac{T}{3{\omega_m}-1}\Big)^{\frac{-3\omega_m}{3(1+\omega_m)}}-\kappa^2\,.
\end{equation}
By integrating the above equation with respect to $T$, we get
\begin{equation}
f_2(T)=-A(69q^2-25q+2){\tau_0}^{-9q}\Big(\frac{T}{3\omega_m-1}\Big)^
{\frac{1}{1+\omega_m}}-\kappa^2T\,.
\end{equation}
Finally, we insert $f(R,T)$ in equation (26) and obtain the
unimodular Lagrange multiplier, $\lambda$, as follows
\begin{equation}
\lambda(\tau)=N_{+}\tau^{\mu_{+}+6q-2}+N_{-}\tau^{\mu_{-}+6q-2}+N_1
\tau^{-3q}+N_2 \tau^{-3(1+\omega_m)q}\,.
\end{equation}
where by definition
\begin{equation}
N_{\pm}=\Big[q(1-6q)+3q(5q-1)\Big(\frac{6q-2}{\mu_{\pm}+6q-2}\Big)+5\mu_{\pm}q+
\mu_{\pm}(\mu_{\pm}-1)\Big]\tau_0^{-6q}C_{\pm}\,,
\end{equation}
\begin{equation}
N_1=-\frac{1}{2}(3\omega_{m}-1)(69q^2-25q+2){{\tau_0}^{-6q}}A\,,
\end{equation}
and
\begin{equation}
N_2=\frac{1}{2}(1-\omega_m)\kappa^2 {\tau_0}^{3(1+\omega_m)q} \,
\end{equation}
We see from Eq. (57) that in this setup the Lagrange multiplier
$\lambda$, which is expected to mimic a cosmological constant in
unimodular viewpoint, is a time varying quantity. This is an
interesting result since an evolving cosmological ``constant"
provides new facilities for the
rest of cosmology, especially for the late time cosmic dynamics.\\

$\bullet \quad f(R,T)=R+2f(T)$ \vspace{0.4cm}

The second specific type of $f(R,T)$ which we consider, is defined
as
\begin{equation}
f(R,T)=R+2f(T) \,.
\end{equation}
In this case, by solving equation (27) we obtain
\begin{equation}
f(T)=\frac{3q^2}{2\tau_0^2}\Big(3\omega-1\Big)^{\frac{3(3+\omega)q-2}{3(1+\omega)q}}
T^{\frac{2(1-3q)}{3(1+\omega)q}}-\frac{1}{2}\kappa^2T\,.
\end{equation}
Then, by substituting this equation into equation (42) and
integrating of it we obtain
\begin{eqnarray}
\lambda(\tau)=-\frac{q}{\tau_0^2}
(\frac{15}{2}q-\frac{9}{2}q\omega_m+2)\big(\frac{\tau}
{\tau_0}\big)^{6q-2}+\frac{1}{2}(1-\omega_m)\kappa^2
\big(\frac{\tau}{\tau_0}\big)^{-3(1+\omega)q}\,.
\end{eqnarray}
So, by using the reconstruction method we were able to obtain an
explicit form of $f(R,T)$ that generates the scale factor (32). Also
we were able to find temporal variation of the unimodular parameter,
$\lambda$.

\section{Cosmological Inflation in Unimodular $f(R,T)$ Gravity}

Now we study cosmological inflation in unimodular $f(R,T)$ gravity. Some important inflation parameters such as the
scalar spectral index, its running, tensor spectral index and
the tensor-to-scalar ratio are obtained in a specific unimodular $f(R,T)$ gravity.
To calculate the inflation parameters, we define the
slow-roll parameters in the Jordan frame as follows~\cite{Mar14}
\begin{eqnarray}
\epsilon_1\equiv-\frac{\dot{H}}{H^2}\,,\hspace{9.25cm}\\
\epsilon_2\equiv\frac{\ddot{H}}{H\dot{H}}-\frac{2\dot{H}}{H^2}\,,\hspace{8.25cm}\\
\epsilon_3\equiv(\ddot{H}H-2\dot{H}^2)^{-1}\Big[\frac{H\dot{H}\dddot{H}-\ddot{H}
(\dot{H}^2+H\ddot{H})}{H\dot{H}}-\frac{2\dot{H}}{H^2}(H\ddot{H}-2\dot{H}^2)\Big]\,.
\end{eqnarray}
Also the inflation parameters are defined as~\cite{Mar14}
\begin{eqnarray}
r\approx16\epsilon_1\,,\hspace{1.7cm}\\
n_s\approx 1-2\epsilon_1-2\epsilon_2\,,\hspace{0.1cm}\\
\alpha_s\approx-2\epsilon_1 \epsilon_2-\epsilon_2 \epsilon_3\,,\\
n_T\approx-2\epsilon_1\,.\hspace{1.45cm}
\end{eqnarray}
To obtain exact forms of the inflation parameters, it is more
convenient to express the slow-roll parameters in terms of the
e-folds number. The e-folds number is defined as
\begin{equation}
N\equiv\ln(\frac{a_f}{a_i})=-\int_{a_i}^{a_f}{H(t)dt}\,,
\end{equation}
where $a_i$ and $a_f$ are the values of the scale factor at the beginning and end of
inflation epoch respectively. In this regard, the slow-roll parameters can be rewritten as
\begin{eqnarray}
\epsilon_1(N)\equiv-\frac{H'(N)}{H(N)}\,,\hspace{10cm}\\
\epsilon_2(N)\equiv\frac{H''(N)}{H'(N)}-\frac{H'(N)}{H(N)}\,,\hspace{8.6cm}\\
\epsilon_3(N)\equiv\Big[\frac{H(N)H'(N)}{H''(N)H(N)-H'^2(N)}\Big]\Big[\frac{H'''(N)}{H'(N)}
-\frac{H''^2(N)}{H'^2(N)}-\frac{H''(N)}{H(N)}+\frac{H'^2(N)}{H^2(N)}\Big]\,.
\end{eqnarray}
Now we study cosmological inflation with power-law scale factor as given by
equation (34). In this case the slow-roll parameters take the following forms
\begin{eqnarray}
\epsilon_1=\frac{1}{\alpha}\,, \qquad \epsilon_2=0\,, \qquad
\epsilon_3=\frac{1}{\alpha}\,,
\end{eqnarray}
leading to the the following inflation parameters
\begin{eqnarray}
r\approx\frac{16}{\alpha}\,, \qquad
n_s\approx1-\frac{2}{\alpha}\,,\qquad \alpha_s\approx0\,, \qquad
n_T\approx-\frac{2}{\alpha}\,.
\end{eqnarray}
In this regard, we obtain the relation between the tensor-to-scalar
ratio and the scalar spectral index as
\begin{equation}
r=8(1-n_s)\,.
\end{equation}
By adopting $\alpha=100$ or $q=0.332$, we obtain $r\approx 0.16$,
$n_s \approx 0.98$, $\alpha_s=0$, and $n_T \approx -0.02$, which are
well in the confidence levels of Planck2015
results~\cite{Ade16,Ade15}. Now, to investigate inflation
dynamics in this setup we consider a specific model of the unimodular $f(R,T)$
gravity which is given by
\begin{equation}
f(R,T)=\alpha_1 R+\alpha_2 T\,,
\end{equation}
where $\alpha_1$ and $\alpha_2$ are constant parameters. Also inspired by equation (6)
 and its counterpart in unimodular $f(R,T)$ gravity we
adopt the following form for the Lagrange multiplier
\begin{equation}
\lambda=\alpha_3 R+\alpha_4 T\,,
\end{equation}
with $\alpha_3$ and $\alpha_4$ being constant parameters. We note
that our motivation to choose $\lambda$ in this form is based on Eq.
(6) in the standard unimodular gravity and also the reconstruction
method as is given by Eq. (63). The assumption for the form of
$\lambda$ in Eq. (79) is compatible with Eq. (78) via the field
equations (17). By substituting these functions into equation (26),
we obtain
\begin{equation}
(6\alpha_3 -2\alpha_1)\dot{H}+(12\alpha_3
-3\alpha_1)H^2=[\kappa^2\omega-\frac{1}{2}(\alpha_{2}-2\alpha_{4})(1-3\omega)]\rho
\,.
\end{equation}
Note that in equation (80) we have used the relation
$d\tau=a^3(\tau)dt$. Differentiating of the above equation with
respect to $t$ gives
\begin{equation}
A\ddot{H}+BH\dot{H}+CH^3=0\,\,,
\end{equation}
where
\begin{eqnarray}
A=6\alpha_3-2\alpha_1 \,\,,\hspace{4.5cm}\nonumber\\
B=2(12\alpha_3-3\alpha_1)+3(1+\omega)(6\alpha_3-2\alpha_1)\,\,,\nonumber\\
C=3(1+\omega)(12\alpha_3-3\alpha_1)\,.\hspace{2.7cm}
\end{eqnarray}
The general solution of the differential equation (81) is obtained
in the parametric form as
\begin{equation}
H=\Big(-\frac{2A}{B}z\Big)^{1/2}\,,
\end{equation}
with
\begin{eqnarray}
z=\zeta \exp \Big(-\int{\frac{\sigma
d\sigma}{\sigma^2-\sigma-D}}\Big)\,\,,\nonumber\\
D=-\frac{2AC}{B^2}\,\,,\hspace{3cm}
\end{eqnarray}
where $\zeta$ is constant and
\begin{eqnarray}
\dot{H}=\sigma z=\sigma \zeta \exp \Big(-\int{\frac{\sigma
d\sigma}{\sigma^2-\sigma-D}}\Big)\,\,,
\end{eqnarray}

\begin{eqnarray}
\ddot{H}=-\frac{B}{A}\Big(\frac{\sigma+D}{\sigma}\Big)H\dot{H}\,.\hspace{2.8cm}
\end{eqnarray}
In this respect, we obtain the e-folds number as
\begin{equation}
N=\frac{2A}{B\sqrt{4D+1}}\,\,\tanh^{-1}\Big(\frac{-2\sigma+1}{\sqrt{4D+1}}\Big)\,\,,
\end{equation}
By using these equations we obtain the slow roll parameters
(64)-(66) as
\begin{eqnarray}
\epsilon_1=\frac{B}{2A}\sigma\,\,,\hspace{2.5cm}\\
\epsilon_2=-\frac{B}{A}\Big(\frac{-\sigma^2+\sigma+D}{\sigma}\Big)\,\,,\\
\epsilon_3=\frac{B}{A}\Big(\frac{\sigma^2+D}{\sigma}\Big)\,\,.\hspace{1.4cm}
\end{eqnarray}
Also, the inflation parameters are obtained as
\begin{eqnarray}
r=8\frac{B}{A}\sigma\,\,,\hspace{4.2cm}\\
n_s=1+2\frac{B}{A}-3\frac{B}{A}\sigma-4\frac{C}{B}\frac{1}{\sigma}\,,\hspace{1cm}\\
\alpha_s=\frac{B^2}{A^2}\Big(-2\sigma^2+2\sigma+\frac{D^2}{\sigma^2}+\frac{D}{\sigma}\Big)\,\,,\\
n_T=-\frac{B}{A}\sigma\,\,,\hspace{4cm}
\end{eqnarray}
We can eliminate $\sigma$ by using equation (91) to find finally
\begin{eqnarray}
n_s=1+\frac{2B}{A}-\frac{3}{8}r-\frac{32C}{A}\frac{1}{r}\,,\hspace{3.6cm}\\
\alpha_s=-\frac{1}{32}r^2+\frac{1}{4}\frac{B}{A}r+\frac{256C^2}{A^2}\frac{1}{r^2}
-\frac{16BC}{A^2}\frac{1}{r}-\frac{2C}{A}\,,\\
n_T=-\frac{1}{8}r\,.\hspace{6.62cm}
\end{eqnarray}

To test the observational viability of our unimodular $f(R,T)$
model, we compare the model with Planck2015 observational data. To
this end, we perform some numerical analysis on the scalar spectral
index and the tensor-to-scala ratio as obtained analytically above. The results are shown in figure 2.
In this figure we have plotted $r$ versus $n_s$ in the background of
the Planck 2015 TT, TE, EE+low P data. To plot this figure, we have
adopted two arbitrary values for $\alpha_{1}$ as $\alpha_{1}=-5$ and
$\alpha_{1}=-10$. With these values of $\alpha_{1}$, the chosen
ranges of $\alpha_2$ are based on the observationally viable values
of $n_s$. We see that this unimodular extension of $f(R,T)$ cosmology is consistent with
Planck2015 observation at least in some subspaces of the model parameter space.

\begin{figure}[htp]
\begin{center}\includegraphics{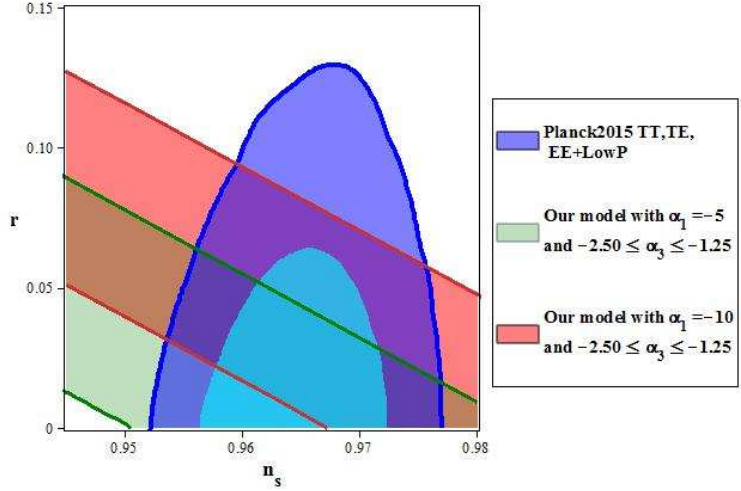} \vspace{7cm}
\end{center}
\caption{\small{ Tensor-to-scalar ratio versus the scalar spectral
index in the background of Planck 2015 TT, TE, EE+low P data for the
specific unimodular $f(R,T)$ as $f(R,T)=\alpha_1 R+\alpha_2 T$ and a
power-law scale factor as given by equation (34).}}
\end{figure}

\section{Unimodular $f(R,T)$ gravity in Einstein frame}

The action of unimodular $f(R,T)$ gravity as given by Eq. (7) corresponds to a
generally nonlinear function $f(R,T)$ in terms of the Ricci scala $R$ and
trace of the stress-energy tensor of the matter, $T_{\mu\nu}$. It is
possible to perform a conformal transformation to Einstein frame in
which the action is conformally equivalent to an Einstein theory
plus a term representing non-minimal coupling to the matter
component. The issue of unimodular modified gravity in Einstein frame
has not been studied thoroughly (except a short discussion in~\cite{Sae16}).
 So, here we switch to the Einstein frame to fill this gap by the following conformal
transformation~\cite{Far98,Fuj03,Mae89,Wan94,Gos13,Vol14}
\begin{equation}
\tilde{g}_{\mu\nu}=\Omega^2 g_{\mu\nu}\,,
\end{equation}
where $\Omega^2$ is the conformal factor and the quantities in
Einstein frame are represented by a tilde. The Christoffel
symbol, Ricci tensor and the Ricci scalar in Einstein frame are
related to the corresponding parameters in Jordan frame as follows
\begin{eqnarray}
\tilde{\Gamma}^{\alpha}_{\mu\nu}=\Gamma^{\alpha}_{\mu\nu}+\Omega^{-1}
\big(\delta^{\alpha}_{\mu}\nabla_{\nu}\Omega+
\delta^{\alpha}_{\nu}\nabla_{\mu}\Omega-g_{\mu\nu}\nabla^{\alpha}\Omega\big)\,,\hspace{7cm}\\
\tilde{R}_{\mu\nu}=R_{\mu\nu}-2\nabla_{\mu}\nabla_{\nu}(\ln\Omega)-g_{\mu\nu}\Box(\ln\Omega)
+2\nabla_{\mu}(\ln\Omega)\nabla_\nu(\ln\Omega)
-2g_{\mu\nu}\nabla^{\alpha}(\ln\Omega)\nabla_\alpha(\ln\Omega)\,,\hspace{0.4cm}\\
\tilde{R}=\Omega^{-2}\Big[R-6\Box(\ln\Omega)-6\nabla^{\alpha}(\ln\Omega)
\nabla_\alpha(\ln\Omega)\Big]\,,\hspace{7.3cm}
\end{eqnarray}
where
$\Box=\Omega^2\tilde{\Box}-2\tilde{\nabla}^{\sigma}(\ln\Omega)\partial_{\sigma}$.
To rewrite action (7) in Einstein frame, we consider $\Omega$ as
\begin{equation}
\Omega^2=\frac{f_{,R}}{1+\frac{f_{,T}}{\kappa^2}}\,,
\end{equation}
and define a new scalar field $\phi$ as
\begin{eqnarray}
\kappa\phi\equiv\sqrt{\frac{3}{2}}\ln\Omega^2=\kappa(\phi_1-\phi_2)\,\,,
\end{eqnarray}
where
\begin{eqnarray}
\kappa\phi_1\equiv\sqrt{\frac{3}{2}}\ln f_{,R} \,\,,\hspace{2.1cm} \\
\kappa\phi_2\equiv\sqrt{\frac{3}{2}}\ln
(1+\frac{f_{,T}}{\kappa^2})\,.\hspace{1.2cm}
\end{eqnarray}
By these definitions, using the relation $\sqrt{-\tilde{g}}=\Omega^4
\sqrt{-g}$, and equations (99)-(101) we reach the following action
written in Einstein frame
\begin{eqnarray}
S=\int
d^4x\bigg\{\sqrt{-\tilde{g}}\bigg(\frac{\tilde{R}}{2\kappa^2}-
\frac{1}{2}\tilde{g}^{\mu\nu}{\nabla_\mu
\phi_1}{\nabla_\nu \phi_1}-\frac{1}{2}\tilde{g}^{\mu\nu}{\nabla_\mu
\phi_2}{\nabla_\nu \phi_2}+\tilde{g}^{\mu\nu}{\nabla_\mu
\phi_1}{\nabla_\nu \phi_2}\hspace{1cm} \nonumber\\
-V(\phi_1,\phi_2)\bigg)-2\tilde{\lambda}
\bigg({\sqrt{-\tilde{g}}}{e^{-2\sqrt{2/3}\kappa(\phi_1
-\phi_2)}-1}\bigg)\bigg\}+\int d^4x
\sqrt{-\tilde{g}}{\cal{L}}_{m}(\Omega^{-2}\tilde{g}_{\mu\nu},\psi_m)\,.
\end{eqnarray}
where $\tilde{\lambda}=\frac{\lambda}{2\kappa^2}$, and
\begin{equation}
V(\phi_1,\phi_2)=\frac{\Omega^2R-f}{2\kappa^2\Omega^4}
\end{equation}
is the potential of the scalar fields  in Einstein frame. This
action contains two canonical scalar fields with a mixed kinetic
term. We can see the effect of the conformal transformation on the
matter Lagrangian in action (106). From equation (103) we have
$\Omega^2=\exp(\sqrt{2/3}\kappa(\phi_1-\phi_2))=F(\phi_1,\phi_2)$.
In this regard, the scalar fields $\phi_1$ and $\phi_2$ are directly
coupled to the matter fields in Einstein frame. Varying action (106)
with respect to the Lagrange multiplier gives
\begin{equation}
\sqrt{-\tilde{g}}=e^{2\sqrt{2/3}\kappa(\phi_1 -\phi_2)}\,.
\end{equation}
Equation (108) shows that, unlike the Jordan frame, in Einstein
frame the determinant of the metric is not a constant quantity
anymore. We can express the Lagrangian density of the fields in the
following form
\begin{eqnarray}
&{\cal{L}_{\phi}}=
-\frac{1}{2}\tilde{g}^{\mu\nu}\nabla_{\mu}\phi_1\nabla_{\nu}\phi_1
-\frac{1}{2}\tilde{g}^{\mu\nu}\nabla_{\mu}\phi_2\nabla_{\nu}\phi_2
+\tilde{g}^{\mu\nu}\nabla_{\mu}\phi_1\nabla_{\nu}\phi_2
-V(\phi_1,\phi_2)-2\tilde{\lambda}e^{-2\sqrt{2/3}\kappa(\phi_1
-\phi_2)}\,.\nonumber \\
\end{eqnarray}
In this regard, the stress-energy tensor is obtained as
\begin{eqnarray}
\tilde{T}_{\mu\nu}(\phi_1,\phi_2)=-\frac{2}{\sqrt{-\tilde{g}}}\frac{\partial
(\sqrt{-\tilde{g}}{\cal{L}_{\phi}})}{\partial\tilde{g}^{\mu\nu}}=-2\frac{\partial
{\cal{L}_{\phi}}}{\partial\tilde{g}^{\mu\nu}}+\tilde{g}_{\mu\nu}
{\cal{L}_{\phi}}\hspace{5.1cm}\nonumber \\
=\nabla_{\mu}\phi_1\nabla_{\nu}\phi_1+\nabla_{\mu}\phi_2\nabla_{\nu}\phi_2
-2\nabla_{\mu}\phi_1\nabla_{\nu}\phi_2-
\tilde{g}_{\mu\nu}\Big[\frac{1}{2}\tilde{g}^{\alpha\beta}\nabla_{\alpha}
\phi_1\nabla_{\beta}\phi_1+\hspace{0.6cm}\nonumber \\
\frac{1}{2}\tilde{g}^{\alpha\beta}\nabla_{\alpha}\phi_2\nabla_{\beta}\phi_2-
\tilde{g}^{\alpha\beta}\nabla_{\alpha}\phi_1\nabla_{\beta}\phi_2+V(\phi_1,\phi_2)\Big]
-2\tilde{\lambda}\tilde{g}_{\mu\nu}e^{-2\sqrt{2/3}\kappa(\phi_1
-\phi_2)}\,.
\end{eqnarray}
By varying the action (106) with respect to the scalar fields
$\phi_1$ and $\phi_2$, we obtain
\begin{eqnarray}
\tilde{\Box} \phi_1-\tilde{\Box}
\phi_2-V_{,\phi_1}+4\tilde{\lambda}\sqrt{\frac{2}{3}}\kappa
e^{-2\sqrt{\frac{2}{3}}\kappa(\phi_1-\phi_2)}+\frac{1}
{\sqrt{-\tilde{g}}}\frac{\partial(\sqrt{-\tilde{g}}
{\cal{L}}_m)}{\partial\phi_1}=0\,,
\end{eqnarray}
\begin{eqnarray}
\tilde{\Box} \phi_2-\tilde{\Box}
\phi_2-V_{,\phi_2}-4\tilde{\lambda}\sqrt{\frac{2}{3}}\kappa
e^{-2\sqrt{\frac{2}{3}}\kappa(\phi_1-\phi_2)}+\frac{1}
{\sqrt{-\tilde{g}}}\frac{\partial(\sqrt{-\tilde{g}}
{\cal{L}}_m)}{\partial\phi_2}=0\,.
\end{eqnarray}
The stress-energy tensor of the matter fields in Einstein frame is
defined as follows
\begin{eqnarray}
\tilde{T}_{\mu\nu}^{(m)}=\frac{-2}{\sqrt{-\tilde{g}}}\frac{\delta(\sqrt{-\tilde{g}}{\cal{L}}_m)}
{\delta\tilde{g}_{\mu\nu}}=F^{-1}T_{\mu\nu}\,.
\end{eqnarray}
Also, we have
\begin{eqnarray}
\frac{\partial(\sqrt{-\tilde{g}}{\cal{L}}_m)}{\partial\phi_1}
=\frac{(\sqrt{-\tilde{g}}{\cal{L}}_m)}{\partial\tilde{g}^{\mu\nu}}\frac{\partial\tilde{g}^{\mu\nu}}{\partial
g^{\mu\nu}}\frac{\partial
g^{\mu\nu}}{\partial\phi_1}=(\frac{-\sqrt{-\tilde{g}}}{2}\tilde{T}_{\mu\nu})
(\Omega^{-2})(\sqrt{\frac{2}{3}}\kappa\Omega^2\tilde{g}^{\mu\nu})=-\frac{\sqrt{-\tilde{g}}}{\sqrt{6}}\kappa
\tilde{T}\,,
\end{eqnarray}
and
\begin{eqnarray}
\frac{\partial(\sqrt{-\tilde{g}}{\cal{L}}_m)}{\partial\phi_2}
=\frac{\sqrt{-\tilde{g}}}{\sqrt{6}}\kappa \tilde{T}\,.
\end{eqnarray}
We show the coupling term between the scalar fields and matter by
$Q$ as
\begin{eqnarray}
Q=-\frac{F_{,\phi_1}}{2\kappa F}=\frac{F_{,\phi_2}}{2\kappa
F}=-\frac{1}{\sqrt{6}}\,,
\end{eqnarray}
where takes a constant value in this model. By substituting
equations (113)-(116) in equations of motion (111) and (112) we find
\begin{eqnarray}
\tilde{{\Box}} \phi_1-\tilde{\Box}
\phi_2-V_{,\phi_1}+4\tilde{\lambda}\sqrt{\frac{2}{3}}\kappa
e^{-2\sqrt{\frac{2}{3}}\kappa(\phi_1-\phi_2)}+\kappa Q
\tilde{T}^{(m)}=0\,,
\end{eqnarray}
\begin{eqnarray}
\tilde{\Box} \phi_2-\tilde{\Box}
\phi_1-V_{,\phi_2}-4\tilde{\lambda}\sqrt{\frac{2}{3}}\kappa
e^{-2\sqrt{\frac{2}{3}}\kappa(\phi_1-\phi_2)}-\kappa Q
\tilde{T}^{(m)}=0\,.
\end{eqnarray}
We obtain the field equations in Einstein frame by varying action
(106) with respect to the metric $\tilde{g}_{\mu\nu}$ as follows
\begin{eqnarray}
\tilde{R}_{\mu\nu}-\frac{1}{2}\tilde{g}_{\mu\nu}\tilde{R}=
\kappa^2(\tilde{T}_{\mu\nu}+\tilde{T}^{(m)}_{\mu\nu})\,,
\end{eqnarray}
where $\tilde{T}_{\mu\nu}$ is the stress-energy of the scalar fields
defined in equation (110). Divergence of the field equations, using
the Bianchi identities
$\nabla^{\mu}(\tilde{R}_{\mu\nu}-\frac{1}{2}\tilde{g}_{\mu\nu}\tilde{R})=0$,
gives
\begin{eqnarray}
\nabla^{\mu}\tilde{T}_{\mu\nu}(\phi_1,\phi_2)=\bigg(\tilde{{\Box}}
\phi_1-\tilde{\Box}
\phi_2-V_{,\phi_1}+4\tilde{\lambda}\sqrt{\frac{2}{3}}\kappa
e^{-2\sqrt{\frac{2}{3}}\kappa(\phi_1-\phi_2)}\bigg)\tilde{\nabla}_{\nu}\phi_1+\hspace{1.5cm}\nonumber\\
\bigg(\tilde{\Box} \phi_2-\tilde{\Box}
\phi_1-V_{,\phi_2}-4\tilde{\lambda}\sqrt{\frac{2}{3}}\kappa
e^{-2\sqrt{\frac{2}{3}}\kappa(\phi_1-\phi_2)}\bigg)\tilde{\nabla}_{\nu}\phi_2
-2e^{-2\sqrt{\frac{2}{3}}\kappa(\phi_1-\phi_2)}\tilde{\nabla_{\nu}}\tilde{\lambda}=0\,.
\end{eqnarray}
Taking into account the conservation of the matter fields,
$\nabla^{\mu}T_{\mu\nu}=0$, in Einstein frame becomes
\begin{eqnarray}
\tilde{\nabla}^{\mu}\tilde{T}_{\mu\nu}=-\frac{1}{\sqrt{6}}\kappa\tilde{T}^{(m)}
\Big(\tilde{\nabla}_{\nu}\phi_1-\tilde{\nabla}_{\nu}\phi_2\Big)
\end{eqnarray}
and by using equations (117), (118) and (121) we get
\begin{eqnarray}
\tilde{\nabla_{\nu}}\tilde{\lambda}=0\,, \quad \rightarrow \quad
\tilde{\lambda}=\tilde{\lambda}_0\,.
\end{eqnarray}
This means that in Einstein frame the unimodularity multiplier $\lambda$ is a constant whereas in
Jordan frame, as we have shown previously, it varies with time. To proceed further, we define the
effective potential as follows
\begin{equation}
V_{eff}(\phi_1,\phi_2)=\frac{FR-f}{2\kappa^2F^2}+2\tilde{\lambda}_0
e^{-2\sqrt{\frac{2}{3}}\kappa(\phi_1-\phi_2)}\,.
\end{equation}
Now, by using the FRW metric in Einstein frame, given by
\begin{eqnarray}
d\tilde{s}^2=\Omega^2
ds^2=-\tilde{a}^{-6}(\tilde{\tau})d\tilde{\tau}^2+\tilde{a}^2(\tilde{\tau})dx_idx^i
\end{eqnarray}
where
\begin{equation}
\tilde{a}=\sqrt{F}a\,\,,\qquad d\tilde{\tau}=F^2 d\tau\,\,,
\end{equation}
we obtain the Friedmann equations in Einstein frame as follows
\begin{eqnarray}
\tilde{{\cal{H}}}^2=\frac{\kappa^2}{3}\bigg[\frac{1}{2}\dot{\phi}_{1}^{2}
+\frac{1}{2}\dot{\phi}_{2}^{2}-\dot{\phi}_{1}\dot{\phi}_{2}+\tilde{a}^{-6}
V_{eff}(\phi_1,\phi_2)+\tilde{a}^{-6}\tilde{\rho}_m\bigg]\,\,,
\end{eqnarray}
\begin{eqnarray}
-2\dot{\tilde{{\cal{H}}}}-9\tilde{{\cal{H}}}^2=\kappa^2\bigg[\frac{1}{2}\dot{\phi}_{1}^{2}
+\frac{1}{2}\dot{\phi}_{2}^{2}-\dot{\phi}_{1}\dot{\phi}_{2}-\tilde{a}^{-6}
V_{eff}(\phi_1,\phi_2)+\tilde{a}^{-6}\tilde{p}_m\bigg]\,,
\end{eqnarray}
where
\begin{eqnarray}
\tilde{{\cal{H}}}\equiv\frac{1}{\tilde{a}}\frac{d \tilde{a}}{d
\tilde{\tau}}=\frac{1}{F^2}\Big[\frac{1}{2F}\frac{d
F}{d\tau}+{\cal{H}}\Big]\,\,.
\end{eqnarray}

To study cosmological inflation in Einstein frame we set ${\cal{L}}_m=0$ and
study the inflationary dynamics in the absence of the matter fields.
We adopt specific type of $f(R,T)$ as
\begin{equation}
f(R,T)=\alpha R+\beta T^n\,\,,
\end{equation}
where $\alpha$, $\beta$ and $n$ are constant. With this definition
of $f(R,T)$, the scalar fields (104) and (105) take the following
form
\begin{eqnarray}
\kappa \phi_1=\sqrt{\frac{3}{2}}\ln{\alpha}=constant \quad , \quad
\kappa \phi_2=\sqrt{\frac{3}{2}}\ln\Big({1+\frac{\beta n
T^{n-1}}{\kappa^2}}\Big)\,\,.
\end{eqnarray}
In this regard, the effective potential becomes
\begin{eqnarray}
V_{eff}(\phi_1,\phi_2)=\frac{1}{2\kappa^2
\alpha^2}e^{2\sqrt{\frac{2}{3}}\kappa
\phi_2}\bigg[\Big(e^{-\sqrt{\frac{2}{3}}\kappa \phi_2}-1\Big)\alpha R-\beta
T^n+2\tilde{\lambda}_0\bigg]\,.
\end{eqnarray}
Under the slow-roll approximations $\dot{\phi}<<V_{eff}$ and
$\ddot{\phi}<<3\tilde{{\cal{H}}}\dot{\phi}$ and by using equations
(126) and (127) we obtain the slow-roll parameters in Einstein frame
as follows
\begin{eqnarray}
\epsilon=\frac{1}{2\kappa^2}\bigg(\frac{V_{eff}'}{V_{eff}}\bigg)
\quad , \quad \eta=\frac{1}{\kappa^2}\bigg(\frac{V_{eff}''}{V_{eff}}\bigg)\,\,.
\end{eqnarray}
By the effective potential as given by equation (131), we get
\begin{eqnarray}
\epsilon= \frac{1}{3}\Bigg[2-\frac{\alpha
e^{-\sqrt{\frac{2}{3}}\kappa
\phi_2}R-\frac{\kappa^2}{n-1}e^{\sqrt{\frac{2}{3}}\kappa
\phi_2}\Big[\frac{\kappa^2}{\beta n}\big(e^{\sqrt{\frac{2}{3}}\kappa
\phi_2}-1\big)\Big]^{\frac{1}{n-1}}}{\big(
e^{-\sqrt{\frac{2}{3}}\kappa \phi_2}-1\big)\alpha
R-\beta\Big[\frac{\kappa^2}{\beta n}\big(e^{\sqrt{\frac{2}{3}}\kappa
\phi_2}-1\big)\Big]^{\frac{n}{n-1}}+2\tilde{\lambda}_0}\Bigg]^2\,\,,
\end{eqnarray}

\begin{eqnarray}
\eta=\frac{2}{3}\Bigg[4-\frac{3\alpha e^{-\sqrt{\frac{2}{3}}\kappa
\phi_2}R-\frac{5\kappa^2}{n-1}e^{\sqrt{\frac{2}{3}}\kappa
\phi_2}\Big[\frac{\kappa^2}{\beta n}\big(e^{\sqrt{\frac{2}{3}}\kappa
\phi_2}-1\big)\Big]^{\frac{1}{n-1}}-\frac{\kappa^2}{\beta n
(n-1)^2}e^{2\sqrt{\frac{2}{3}}\kappa
\phi_2}\Big[\frac{\kappa^2}{\beta n}\big(e^{\sqrt{\frac{2}{3}}\kappa
\phi_2}-1\big)\Big]^{\frac{2-n}{n-1}}}{\big(e^{-\sqrt{\frac{2}{3}}\kappa
\phi_2}-1\big)\alpha R-\beta\Big[\frac{\kappa^2}{\beta
n}\big(e^{\sqrt{\frac{2}{3}}\kappa
\phi_2}-1\big)\Big]^{\frac{n}{n-1}}+2\tilde{\lambda}_0}\Bigg]\,\,,\nonumber
\\
\end{eqnarray}
In this regard, we obtain other inflation parameters in Einstein
frame as
\begin{eqnarray}
r=16\epsilon= \frac{16}{3}\Bigg[2-\frac{\alpha
e^{-\sqrt{\frac{2}{3}}\kappa
\phi_2}R-\frac{\kappa^2}{n-1}e^{\sqrt{\frac{2}{3}}\kappa
\phi_2}\Big[\frac{\kappa^2}{\beta n}\big(e^{\sqrt{\frac{2}{3}}\kappa
\phi_2}-1\big)\Big]^{\frac{1}{n-1}}}{\big(e^{-\sqrt{\frac{2}{3}}\kappa
\phi_2}-1\big)\alpha R-\beta\Big[\frac{\kappa^2}{\beta
n}\big(e^{\sqrt{\frac{2}{3}}\kappa
\phi_2}-1\big)\Big]^{\frac{n}{n-1}}+2\tilde{\lambda}_0}\Bigg]^2\,\,,
\end{eqnarray}

\begin{eqnarray}
n_s=\frac{-1}{3}\Bigg[5-\frac{12\alpha e^{-\sqrt{\frac{2}{3}}\kappa
\phi_2}R-\frac{8\kappa^2}{n-1}e^{\sqrt{\frac{2}{3}}\kappa
\phi_2}\Big[\frac{\kappa^2}{\beta n}\big(e^{\sqrt{\frac{2}{3}}\kappa
\phi_2}-1\big)\Big]^{\frac{1}{n-1}}+\frac{4\kappa^2}{\beta n
(n-1)^2}e^{2\sqrt{\frac{2}{3}}\kappa
\phi_2}\Big[\frac{\kappa^2}{\beta n}\big(e^{\sqrt{\frac{2}{3}}\kappa
\phi_2}-1\big)\Big]^{\frac{2-n}{n-1}}}{\big(e^{-\sqrt{\frac{2}{3}}\kappa
\phi_2}-1\big)\alpha R-\beta\Big[\frac{\kappa^2}{\beta
n}\big(e^{\sqrt{\frac{2}{3}}\kappa
\phi_2}-1\big)\Big]^{\frac{n}{n-1}}+2\tilde{\lambda}_0}\Bigg]\nonumber
\\
-2\Bigg[\frac{\alpha e^{-\sqrt{\frac{2}{3}}\kappa
\phi_2}R-\frac{\kappa^2}{n-1}e^{\sqrt{\frac{2}{3}}\kappa
\phi_2}\Big[\frac{\kappa^2}{\beta n}\big(e^{\sqrt{\frac{2}{3}}\kappa
\phi_2}-1\big)\Big]^{\frac{1}{n-1}}}{\big(e^{-\sqrt{\frac{2}{3}}\kappa
\phi_2}-1\big)\alpha R-\beta\Big[\frac{\kappa^2}{\beta
n}\big(e^{\sqrt{\frac{2}{3}}\kappa
\phi_2}-1\big)\Big]^{\frac{n}{n-1}}+2\tilde{\lambda}_0}\Bigg]^2\hspace{6.3cm}
\end{eqnarray}
and
\begin{eqnarray}
n_T= -\frac{2}{3}\Bigg[2-\frac{\alpha e^{-\sqrt{\frac{2}{3}}\kappa
\phi_2}R-\frac{\kappa^2}{n-1}e^{\sqrt{\frac{2}{3}}\kappa
\phi_2}\Big[\frac{\kappa^2}{\beta n}\big(e^{\sqrt{\frac{2}{3}}\kappa
\phi_2}-1\big)\Big]^{\frac{1}{n-1}}}{\big(e^{-\sqrt{\frac{2}{3}}\kappa
\phi_2}-1\big)\alpha R-\beta\Big[\frac{\kappa^2}{\beta
n}\big(e^{\sqrt{\frac{2}{3}}\kappa
\phi_2}-1\big)\Big]^{\frac{n}{n-1}}+2\tilde{\lambda}_0}\Bigg]^2\,\,,
\end{eqnarray}
To see the observational viability of the model in Einstein frame,
we perform a numerical analysis on the scalar spectral index and
tensor-to-scala ratio and compare the results with Planck2015 data.
The results are shown in figure 3. To plot this figure we have
adopted $-0.79<\beta<-0.76$ based on the observationally viable
values of the scalar spectral index. As figure shows, this model in
Einstein frame for some values of the parameters is consistent with
observational data.

\begin{figure}[htp]
\begin{center}\includegraphics{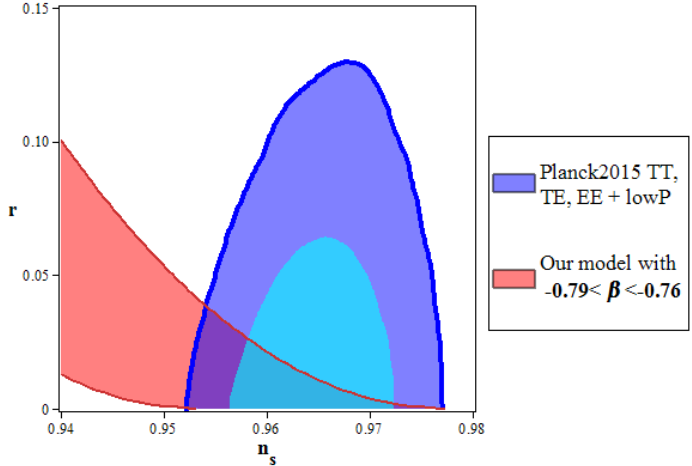} \vspace{6cm}
\end{center}
\caption{\small{Tensor-to-scalar ratio versus the scalar spectral
index in the background of Planck 2015 TT, TE, EE+low P data for the
specific unimodular $f(R,T)$ as $f(R,T)=\alpha R+\beta T^n$.}}
\end{figure}

At this point we note that the issue of frames and their possible
equivalence is an important issue in general relativity and
cosmology (see for instance
Refs~\cite{Vol14,Fara07,Noz09,Domen15,Kamens15}. There are some
controversies on the equivalence of these two frames in quantum and
even classical levels. Here we have done our analysis in two frames
separately in order to see the situation from a modified unimodular
gravity perspective. We noticed that the Lagrange multiplier works
differently in these two frames. This multiplier depends on cosmic
time in Jordan frame and therefore it can act as an evolving scalar field in
the universe evolution. However, in the Einstein frame it acts as a
cosmological constant. So, from a  modified unimodular gravity
perspective these two frames are not equivalent, at least on this
ground.

\section{Conclusion}
Unimodular gravity provides a simple mathematical framework for realization of a cosmological constant.
From a classical viewpoint, unimodular gravity recovers the same
physics as the standard general relativity with a cosmological constant. However, the cosmological
constant appears in unimodular gravity as an integration constant (or a Lagrange multiplier) which can take
any value. While modified theories of gravity, such as $f(R)$ gravity,
are essentially capable to realize a cosmological constant,
the motivation to introduce unimodular modified gravity is to reveal
some new features and cosmological solutions that were impossible to be
realized in standard modified theories of gravity. For instance, it has been
shown in Ref.~\cite{Noj16} that within the framework of reconstruction method it is
possible to realize various cosmological solutions which were impossible in standard unimodular
gravity. In this paper we have extended the idea of unimodular gravity to the modified $f(R,T)$
gravities in order to shed light on some yet unknown cosmological solutions in the spirit of unimodular modified gravities. Our
main motivation was to explore some yet unknown aspects of cosmological solutions in the spirit of
$f(R,T)$ theories. We have shown that a unimodular extension of $f(R,T)$ theories has the potential to reveal
some new features that were impossible to be realized in standard $f(R,T)$ theories.
By introducing a new time variable, we derived the field equations in Jordan frame at the first step.
We have shown that in the Jordan frame the Lagrange multiplier, that imposes the unimodularity constraint on the action,
depends on the newly defined cosmic time. This means that this multiplier is capable to act as a scalar
field in the cosmic evolution. This feature potentially sheds light on the dark energy problem and
can lead to the late time cosmic accelerated expansion. Then we have transformed to the
Einstein frame where the Lagrange multiplier, unlike the Jordan frame, now is a
constant. As we have shown, unlike the Jordan frame, the determinant of the
metric in the Einstein frame is not a constant. We have used the standard reconstruction
method to find some explicit form of unimodular $f(R,T)$ corresponding
to a given cosmological solution. In this paper, we were able to give
some explicit time dependent form for unimodular Lagrange multiplier in Jordan frame.
Since the lagrange multiplier in this setup is expected to mimic a
cosmological constant in unimodular viewpoint, this time varying feature is an interesting
result since an evolving cosmological "constant" provides new facilities for the rest of cosmology,
especially for the late time cosmic speed up. We studied cosmological inflation in this
unimodular $f(R,T)$ scenario both in Jordan and Einstein frames. Investigation of tensor-to-scalar
ratio versus the scalar spectral index shows that there are subspaces of the model parameter space
that the model is consistent with PLANCK2015 observational data. Finally we note that
cosmological perturbations of the comoving curvature perturbation, originating from primordial quantum fluctuations in $f(R,T)$
unimodular gravity are expected to be the same as the one in ordinary $f(R,T)$ gravity at least in linear
perturbation level.

In summary, after constructing a \emph{unimodular} $f(R,T)$ gravity for the first time, we have studied the cosmological dynamics in this setup.
We were able to find some new solutions that are impossible to be realized in \emph{standard} $f(R,T)$ theory.
After reconstruction of an explicit unimodular $f(R,T)$, we have shown that late time cosmic speed up and initial cosmological inflation can be realized in this framework successfully.
Our numerical study of the model parameter space and confrontation with observational data of Planck2015 shows consistency of this model with observation. We have compared our results in Einstein and Jordan frames and we observed that these two frames are not equivalent in this ground in the sense that the Lagrange multiplier works differently in these two frames. While this multiplier depends on cosmic time in Jordan frame and therefore it can act as an evolving scalar field in the universe history, in the Einstein frame it acts as a cosmological constant.\\

{\bf Acknowledgement}
We would like to thank the referee for very insightful comments that improved the quality of the paper considerably.

\end{document}